# Electronic Structures of Ce$M_2$Al$_{10}$ ($M$ = Fe, Ru, and Os) Studied by Soft X-ray Resonant and High-Resolution Photoemission Spectroscopies

Toshihiko Ishiga[1], Takanori Wakita[1,2], Rikiya Yoshida[1,*], Hiroyuki Okazaki[1,†], Koji Tsubota[1], Masanori Sunagawa[1], Kanta Uenaka[1], Kozo Okada[1], Hiroshi Kumigashira[3], Masaharu Oshima[4,‡], Keisuke Yutani[5], Yuji Muro[6], Toshiro Takabatake[5,7], Yuji Muraoka[1,2], and Takayoshi Yokoya[1,2,§]

[1]The Graduate School of Nature Science and Technology, Okayama University, Okayama 700-8530, Japan

[2]Research Laboratory for Surface Science, Okayama University, Okayama 700-8530, Japan

[3]KEK, Photon Factory, Tsukuba, Ibaraki 305-0801, Japan

[4]Department of Applied Chemistry, University of Tokyo, Bunkyo, Tokyo 113-8656, Japan

[5]Department of Quantum Matter, ADSM, Hiroshima University, Higashi-Hiroshima, Hiroshima 739-8530, Japan

[6]Department of Liberal Arts and Sciences, Toyama Prefectural University, Izumi, Toyama 939-0398, Japan

[7]Institute for Advanced Materials Research, Hiroshima University, Higashi-Hiroshima, Hiroshima 739-8530, Japan

*Present Address:  Research Institute for Electronic Science, Hokkaido University, Sapporo, Hokkaido, 001-0021, Japan

†Present Address:  WPI-Advanced Institute for Materials Research,  Tohoku University, Sendai, Miyagi 980-8577, Japan

‡Present address: Synchrotron Radiation Research Organization, The University of Tokyo, Tokyo 113-8656, Japan

(Dated: May 3, 2014)

Abstract

We have performed a photoemission spectroscopy (PES) study of Ce$M_2$Al$_{10}$ ($M$ = Fe, Ru, and Os) to directly observe the electronic structure involved in the unusual magnetic ordering. Soft X-ray resonant (SXR) PES provides spectroscopic evidence of the hybridization between conduction and Ce 4$f$ electrons (*c-f* hybridization) and the order of the hybridization strength (Ru < Os < Fe). High-resolution (HR) PES of CeRu$_2$Al$_{10}$ and CeOs$_2$Al$_{10}$, as compared with that of CeFe$_2$Al$_{10}$, identifies two structures that can be ascribed to structures induced by the *c-f* hybridization and the antiferromagnetic ordering, respectively. Although the *c-f* hybridization-induced structure is a depletion of the spectral intensity (pseudogap) around the Fermi level ($E_F$) with an energy scale of 20−30 meV, the structure related to the antiferromagnetic ordering is observed as a shoulder at approximately 10−11 meV within the pseudogap. The energies of the shoulder structures of CeRu$_2$Al$_{10}$ and CeOs$_2$Al$_{10}$ are approximately half of the optical gap (20 meV), indicating that $E_F$ is located at the midpoint of the gap.

1. **Introduction**

$f$ electron systems exhibit interesting physical properties, such as magnetism, heavy-fermion behavior, exotic superconductivity, and quantum critical behavior,[1] arising from the interplay between the Kondo effect and Ruderman-Kittel-Kasuya-Yosida (RKKY) interactions.[2] Among them, some compounds showing semiconducting behavior at low temperatures, such as Ce$_3$Bi$_4$Pt$_3$ (Ref. 3) and YbB$_{12}$ (Ref. 4), are called Kondo insulators or Kondo semiconductors. The behavior is known to originate from the energy gap formation at the Fermi level ($E_F$) owing to the hybridization between conduction bands and localized $f$ states (*c-f* hybridization).[5,6]

Ce$M_2$Al$_{10}$ ($M$ = Fe, Ru, and Os), which crystallizes in an orthorhombic YbFe$_2$Al$_{10}$-type structure with the space group *Cmcm*,[7] is a new family of Kondo semiconductors exhibiting anomalous phase transitions at $T_0$ = 27.3 and 28.7 K in CeRu$_2$Al$_{10}$ and CeOs$_2$Al$_{10}$, respectively.[8-10] Although the antiferromagnetic ordering of 4$f$ electrons with a reduced moment of 0.3 − 0.4 μ$_B$ has been confirmed to be the origin of the transition, the mechanism of the magnetic ordering at such a high temperature has been unexplained, in spite of extensive research studies.[11-52] This is because, for the Ce-Ce distance of CeRu$_2$Al$_{10}$ and CeOs$_2$Al$_{10}$ (~ 5 Å) and also for the reduced Ce moment, the transition temperatures are considerably higher than those expected from the RKKY interaction between localized 4$f$ moments.

To understand the mechanism of the transition, the investigation of the electronic structure near $E_F$ is crucial. Resistivity, magnetic susceptibility, specific heat, and NMR/NQR studies have provided a sign of energy gap formation near $E_F$ due to the *c-f* hybridization and the phase transition.[8-12,23,25,28,32,33,45,52] Moreover, polarized optical conductivity measurements[24,30,36] have clearly shown two electronic structures; shoulder structures at photon energies of 35 − 55 meV for three compounds and peak structures at a photon energy of ~ 20 meV only for *E // b* below 32 and 39 K in $CeRu_2Al_{10}$ and $CeOs_2Al_{10}$, respectively. Although the shoulder structures were attributed to the direct transition between bonding and antibonding bands due to the *c-f* hybridization, the peak structures were attributed to the formation of charge density waves (CDWs). However, some of the physical properties may not be easily explained by the formation of CDWs along the *b*-axis.[33] The calculated Fermi surface of $CeRu_2Al_{10}$ has no clear nested sheets along the *b*-axis.[44] To have a thorough understanding of electronic structures, photoemission spectroscopy (PES), which enables the direct observation of occupied electronic states, is valuable and can give complementary results, as optical spectroscopy enables the observation of the joint density of states. The relationship of the energy scales of electronic structures with those of spin gaps observed from inelastic neutron scattering studies[18,25,31,43,51] may give some clues to understanding the mechanism of the phase transition. In addition, although spectroscopic studies of the *c-f* hybridization of $CeM_2Al_{10}$ have been performed using core-level PES (Refs. 36 and 40) and X-ray absorption spectroscopy (XAS) (Refs. 41 and 46), resonant PES can give more direct spectroscopic evidence of the involvement of states near $E_F$ in the *c-f* hybridization.

In this article, we report the results of soft X-ray resonant (SXR) PES and high-resolution (HR) PES of $CeM_2Al_{10}$, which have been performed to directly study the *c-f* hybridization and electronic structures. SXRPES confirms the order of the hybridization strength ($T$ = Ru < Os < Fe). From the temperature-dependent HRPES and careful analyses, in spite of the smaller contribution of the transition-induced electronic structural change expected from the direction dependence as well as the coexistence of the two phenomena, we identified two structures that can be ascribed to the structure induced by the *c-f* hybridization and the phase transition for $CeRu_2Al_{10}$ and $CeOs_2Al_{10}$. This suggests the role of electronic structures in the anomalous magnetic phase transition.

## 2. Experimental procedure

Single crystals of Ce$M_2$Al$_{10}$ ($M$ = Fe, Ru, and Os) were grown using a self-flux method in an alumina crucible sealed in a quartz tube under an Ar atmosphere of 1/3 atm. Details of this method were reported previously.[19]

SXRPES measurements were performed at BL2C, KEK-PF, with an SES2000 electron analyzer (total energy resolution of ~ 250 meV) and linear polarized light. The samples were fractured under a base pressure of better than $1.5 \times 10^{-10}$ Torr to obtain clean surfaces at 20 K and kept under the same conditions during the measurements. X-ray absorption spectroscopy (XAS) measurements with a total electron yield mode were also performed for the same samples.

HRPES studies were performed using a GMMADATA-SCIENTA R4000 electron analyzer with a monochromatic Xe I (8.44 eV) resonance line at Okayama University. The total energy resolution was set to 3.7 meV. The base pressure of the spectrometer was better than $6.0 \times 10^{-9}$ Pa. The position of $E_F$ of the samples was referenced to that of a Au film evaporated near the sample holder, and determined within an accuracy of ± 0.10 meV. Clean surfaces were obtained *in situ* by fracturing the samples under ultrahigh vacuum. The sample temperature was measured with a Pt resistive sensor mounted close to the sample. At higher temperatures, we observed a change in the spectral shape in the higher-binding-energy region, which prevented us from performing reliable PES studies. We therefore use the data that showed no spectral changes in the higher-binding-energy region. This limits the highest measured temperatures (120 K for CeFe$_2$Al$_{10}$, and 70 K for CeRu$_2$Al$_{10}$ and CeOs$_2$Al$_{10}$). Several PES measurements taken at different machine times confirmed the reproducibility of the results. No angular dependences were observed for all the PES spectra.

## 3. Results and discussion

We start with the valence band electronic structure that gives insight into the *c-f* hybridization leading to Kondo semiconducting behavior. Figures 1(a) and 1(b) show the on and off resonant photoemission spectra across the Ce 3*d*-4*f* threshold of Ce$M_2$Al$_{10}$ ($M$ = Fe, Ru, and Os) measured at two photon energies, which are indicated in the XAS spectra [Fig. 1(c)]. The on-resonance spectra of three samples have dominant peaks at $E_F$ and structures at approximately 2eV. On-resonance spectra are dominated by the Ce 4$f$ partial density of states (DOS) due to the resonant enhancement, and the structures at approximately $E_F$ and 2 eV binding energy for all the samples are ascribed to the $f^1$ and $f^0$ final states, respectively.[53] Marked increase in the intensity at $E_F$, compared with the off-resonance spectra, provides direct spectroscopic evidence of the $c$-$f$ hybridization of the states near $E_F$ in Ce$M_2$Al$_{10}$. It is known that the ratio of the $f^1$ final state intensity to the $f^0$ final state intensity ($f^1/f^0$) for the on-resonance spectra reflects the hybridization strength.[54] In the observed spectra, the $f^1/f^0$ ratios of $M$ = Fe, Ru, and Os are estimated to be 4.2, 2.4, and 3.6, respectively. This indicates that the hybridization strength increases in the order of CeRu$_2$Al$_{10}$ < CeOs$_2$Al$_{10}$ < CeFe$_2$Al$_{10}$. The order of the hybridization strength is consistent with that derived from previous studies of bulk properties and optical conductivity.[8,10,36,46]

The order of the hybridization strength obtained from experimental studies is different from the order in the periodic table, *i.e.*, Fe, Ru, and Os. A comparison of the off-resonance spectra [Fig. 1(b)] gives some hints, as discussed using the results of band calculations.[30] Taking into account the photoionization cross sections,[55] the off-resonance spectra are expected to be dominated by transition-metal $d$ electron partial density of states (partial DOS). Therefore, the change in the spectral shape reflects the changes in the shape and location of $d$-derived DOS. The spectrum of CeFe$_2$Al$_{10}$ has a prominent peak at approximately 0.7 eV. That of CeRu$_2$Al$_{10}$ also has a prominent peak at approximately 2.2 eV with a wider bandwidth than that of CeFe$_2$Al$_{10}$. The spectral shape is similar to the shape of the recently reported X-ray PES spectrum of CeRu$_2$Al$_{10}$.[40] In CeOs$_2$Al$_{10}$, the spectrum shows a two-peak structure at binding energies of approximately 1.2 and 3.0 eV with a small structure at approximately 5.5 eV, although the centers of mass of the valence bands of CeRu$_2$Al$_{10}$ and CeOs$_2$Al$_{10}$ remain at nearly the same energy. The two-peak structure is most probably due to Os 5$d$ spin-orbit interaction. This makes the DOS near the $E_F$ region of CeOs$_2$Al$_{10}$ comparable to or even larger than that of CeRu$_2$Al$_{10}$, giving rise to a higher hybridization strength of CeOs$_2$Al$_{10}$ than of CeRu$_2$Al$_{10}$.

On the basis of the spectroscopic confirmation of the $c$-$f$ hybridization, we discuss the electronic structure near $E_F$. Figures 2(a)−2(c) show the temperature dependences of the HRPES spectra near $E_F$ of Ce$M_2$Al$_{10}$ ($M$ = Fe, Ru, and Os, respectively) taken using the Xe I (8.44 eV) resonance line. The spectral intensity is normalized to the spectral area under the curve between the binding energies of -100 and 100 meV. The spectral intensity near $E_F$ for Ce$M_2$Al$_{10}$ decreases toward $E_F$, which is consistent with the spectral shapes of the off-resonance spectra shown above. The temperature-dependent spectral changes exhibit asymmetric temperature dependences with respect to $E_F$, differently from the symmetric variation in intensity observed for Au (not shown). Near $E_F$ (inset), although the spectral intensity at $E_F$ of Au does not change upon cooling, that of Ce$M_2$Al$_{10}$ gradually decreases, which suggests pseudogap formation in Ce$M_2$Al$_{10}$, as observed for other Kondo semiconductors.[56,57] Correspondingly, the temperature-dependent intensities at $E_F$ for the three samples decrease as the temperature decreases [Figs. 2(d)−2(f)]. Compared with the temperature dependence of CeFe$_2$Al$_{10}$, those of CeRu$_2$Al$_{10}$ and CeOs$_2$Al$_{10}$ seem to show a steep reduction at lower temperatures. However, the temperature dependences normalized by the Kondo temperature $T_K$ within the measured temperature regions of CeRu$_2$Al$_{10}$ and CeOs$_2$Al$_{10}$ nearly follow that of CeFe$_2$Al$_{10}$. Here, $T_K$ is estimated to be 3 x $T_{max}$, where $T_{max}$ is the temperature at a broad maximum in the magnetic susceptibility. This suggests that this temperature-dependent reduction is mainly governed by the Kondo semiconducting behavior, and that the portion of the electronic structure responsible for the transition is small, consistent with the direction dependence of the phase transition reported from optical studies.[24,36] Nonetheless, the lowest temperature spectra were noticeably different between CeFe$_2$Al$_{10}$ and compounds exhibiting the phase transitions (CeRu$_2$Al$_{10}$ and CeOs$_2$Al$_{10}$): the spectra at 10 K of CeRu$_2$Al$_{10}$ and CeOs$_2$Al$_{10}$ appear to have a rounder edge than that of CeFe$_2$Al$_{10}$. This observation suggests the correspondence of the electronic structure to the phase transition.

To understand the suppression of the spectral intensity at $E_F$ and the spectral difference at the lowest temperature, we derived the spectral density of states (DOS),[56] as shown in Figs. 2(g)−2(i). The temperature-dependent PES spectra were divided by a Gaussian-broadened ($\Delta E$ = 3.7 meV) Fermi-Dirac(FD) distribution function. Obtained spectra were further divided by FD-divided spectrum at 120 K for CeFe$_2$Al$_{10}$ and by FD-divided spectra at 70 K for CeRu$_2$Al$_{10}$ and CeOs$_2$Al$_{10}$. For CeFe$_2$Al$_{10}$, as the temperature decreases, the normalized DOS up to ~ 30 meV from $E_F$ deceases. The normalized DOSs of CeRu$_2$Al$_{10}$ and CeOs$_2$Al$_{10}$ also show a temperature-dependent suppression of intensity the same as that of CeFe$_2$Al$_{10}$, with energy scales corresponding to this suppression depending on the compound (thick gray lines). In addition, at the lowest measured temperature of 10 K, the DOSs show shoulder structures near $E_F$ (red bars). These indicate two characteristic energy scales for the three compounds.

For quantitative discussion, we determined the binding energies of the two features as the crossing points of lines from line fittings for the higher, middle, and lower energy regions, as shown in Fig. 3. The binding energies where the depression of DOS starts can be estimated to be 30+/-3, 21+/-3, 27+/-6 meV for $M$ = Fe, Ru, and Os, respectively. The energies of the shoulder structures were also determined by the same procedure using fitting lines for the middle- and lower-binding-energy regions (6+/-1, 10+/-2, and 11+/-2 meV for $M$ = Fe, Ru, and Os, respectively). The order of the energies of the higher-binding-energy structures shows good correspondence with that of the $f^1/f^0$ ratios, indicating the close correlation of the higher-binding-energy structure with the $c$-$f$ hybridization. These binding energies of the higher-binding-energy structures are in good agreement with half of the optical gaps (55, 35, and 45 meV for $M$ = Fe, Ru, and Os, respectively), which were attributed to the $c$-$f$ hybridization gap in the literature.[24,30,36] According to the $c$-$f$ hybridization model, as the temperature decreases across $T_K$, the $c$-$f$ hybridization between the highly dispersive conduction band and the nondispersive $f$ level gradually rearranges, resulting in bonding and antibonding bands below and above $E_F$, respectively, with the dispersive part similar to the conduction band at a higher binding energy. Near $E_F$, the heavy $f$-derived bands are located at $\Delta_f$ both sides of $E_F$ at different momentum regions (inset of Fig. 3), while another energy scale, $\Delta_t$, is observed from transport measurements.[56,58] Within the energy region where the conduction band hybridizes with the $f$ level, the conduction band loses its intensity and forms a pseudogap in the conduction-electron-derived DOS within $\Delta_c$.[56] We attribute the higher-binding-energy structures observed for the three compounds to the pseudogap of the conduction band $\Delta_c$. The Kondo scaling of the pseudogap observed from PES was reported for several Ce-based Kondo semiconductors, where the universal relation $\Delta_c \sim 2k_B T_K$ holds for most of the compounds.[57] In Ce$M_2$Al$_{10}$, the estimated $\Delta_c/k_B T_K$ values are 1.7, 2.3, and 2.3 for $M$ = Fe, Ru, and Os, which approximately follow the universal relation.

For the lower-binding-energy shoulder, we first discuss that of 6 meV observed in CeFe$_2$Al$_{10}$, which is a paramagnetic Kondo semiconductor. The temperature dependences of the spin-lattice relaxation rate $1/T_1$ and Knight shift are explained by the opening of the pseudogap, which has a magnitude of 55−70 K (4.7−6.0 meV), on the basis of a model of rectangular DOS.[11,23] In the magnetic contribution of the specific heat divided by the temperature, a Schottky-type peak appears at approximately 30 K, suggesting a gap of 60 K (5.2 meV) between the excited and ground states.[8] The energy of the shoulder structure of 6 meV obtained by HRPES is in good agreement with the values from NMR/NQR and specific heat studies, but larger than the activation energy [15 K (1.3 meV)] obtained by the resistivity measurement,[8] corresponding to $\Delta_t$. Therefore, we attribute the shoulder structure to the $\Delta_f$-related structure that leads to an indirect gap of 12 meV (see the inset of Fig. 3). Inelastic neutron scattering studies reported a spin gap of 13 meV for $CeFe_2Al_{10}$.[48] The ratio of the spin gap to the charge gap is 0.9, which is in line with the theoretical prediction (0.7-0.9) for the Anderson lattice model with infinite dimension.[6,58] The 6 meV structure was not determined by the optical spectroscopy studies, most probably owing to the low intensity of the optically inhibited transition between indirect gaps.

If $\Delta_f$ is proportional to the *c-f* hybridization strength in the three compounds, the energies of the shoulder structures of the *M* = Ru and Os compounds are expected to be smaller than that of the *M* = Fe compound. However, our observations are opposite to this expectation. This suggests that the shoulder structures have a different origin from the *c-f* hybridization. The PES values of the shoulder structures are in good agreement with the activation energy of ~ 120 K (~ 10 meV) below $T_0$s obtained in the NMR and specific heat studies.[28,38] The remaining DOS within the transition-induced gap reported in previous NMR studies for *M* = Os is also consistent with the result of the present study, showing the remaining spectral intensity at $E_F$ and 10 K. Therefore, we attribute the shoulder structures of 10 meV in $CeRu_2Al_{10}$ and 11 meV in $CeOs_2Al_{10}$ to the characteristic electronic structure derived from the phase transition. On the other hand, the activation energies obtained by resistivity measurements above $T_0$ are 30−83 K (2.6−7.2 meV) for *M* = Os (Refs. 38 and 52) and 20−40 K (1.7−3.4 meV) for *M* = Ru (Refs. 9 and 10), which are smaller than those obtained below $T_0$ and are most probably induced by the *c-f* hybridization. The disagreement between the 10−11 meV structure obtained in the present PES studies and the activation energies is not clear at the present stage. Inelastic neutron scattering studies have shown the spin gap energies of the Ru and Os compounds to be 8 and 11 meV, respectively,[18,25] which are comparable to the PES values and may indicate the relation between charge gap and spin gap in the transition from the experimental side. Importantly, the value of the structure (10−11 meV) is in good agreement with half of the energy of the induced structure along the *b*-axis in the polarized optical conductivity measurements (20 meV) for CeRu$_2$Al$_{10}$ and CeOs$_2$Al$_{10}$.[24,36)] Since optical spectroscopy measures only the joint density of states, the gaps determined by PES being approximately half of the optical gaps indicate that $E_F$ is located at the center of the gap.

From their optical spectroscopic studies, Kimura *et al.* proposed that the formation of the CDW along the *b*-axis triggers the magnetic ordering.[36)] The results of the present PES studies, compared with optical studies, suggest that the charge gap related to the phase transition shows a particle-hole symmetry. This is in line with CDW formation, although the calculated Fermi surface sheets of CeRu$_2$Al$_{10}$ do not exhibit nesting behavior along the *b*-axis.[44)] On the other hand, recent studies of electron- and hole-doped CeOs$_2$Al$_{10}$ showed the relationship between the hybridization gap above $T_0$ and the magnetic ordering.[52)] We plan to perform systematic HRPES studies on the doped samples, which can directly observe the relationship of structures due to *c-f* hybridization and the phase transition.

**4. Conclusions**

We have performed SXRPES and HRPES of Ce$T_2$Al$_{10}$ (*M* = Fe, Ru, and Os). SXRPES confirmed *c-f* hybridization and the order of the hybridization strength (Ru < Os < Fe). The results of temperature-dependent HRPES showed pseudogap formation with the onset of the depletion of spectral DOS at ~ 30 meV in CeFe$_2$Al$_{10}$, ~ 20 meV in CeRu$_2$Al$_{10}$, and ~ 20−30 meV in CeOs$_2$Al$_{10}$, which can be ascribed to the pseudogap formed owing to the *c-f* hybridization. At the lowest measured temperature of 10 K, we observe a shoulder structure in the normalized DOS of the three compounds, but with different spectral shapes and energies between the paramagnetic CeFe$_2$Al$_{10}$ and the two magnetically ordered compounds, enabling us to attribute the shoulder structures observed in *M* = Ru and Os to structures induced by the antiferromagnetic phase transition. The energies of the phase-transition-induced structures of CeRu$_2$Al$_{10}$ and CeOs$_2$Al$_{10}$ (10−11 meV) are half of the optical gap (20 meV), indicating that $E_F$ is located at the center of the gap. This suggests a particle-hole symmetry of the gap, which limits the mechanism of the magnetic transition.

**Acknowledgements**

We thank S. Kimura and M. Sera for valuable discussion. We also thank Y. Nakamura, T. Jabuchi, and J. Sonoyama for their assistance in the measurements at PF. The PES measurements at BL2C of PF were performed under the project 2011S2-003. The preliminary PES measurements at BL25SU and BL27SU of SPring-8 were performed under proposal numbers 2009B1705 and 2009B1757. This study was supported by Grants-in-Aid for Scientific Research on Innovative Areas "Heavy Electrons" (Nos. 20102003 and 20102004) from the Ministry of Education, Culture, Sports, Science, and Technology of Japan (MEXT) and for Scientific Research (C) (No. 226400363) Japan Society for the Promotion of Science (JPSJ).

§E-mail: yokoya@cc.okayama-u.ac.jp

1) P. Gegenwart, Q. Si, and F. Steglich, Nat. Phys. **4**, 186 (2008).

2) S. Doniach, Physica B + C **91**, 231 (1977).

3) M. F. Hundly, Phys. Rev. B **42**, 6842 (1990).

4) M. Kasaya, J. Magn. Magn. Mater. **31-34**, 437 (1983).

5) T. Takabatake, F. Iga, T. Yoshino, Y. Echizen, K. Katoh, K. Kobayashi, M. Higa, N. Shimizu, Y. Bando, G. Nakamoto, H. Fujii. K. Izawa, T. Suzuki, T. Fujita, M. Sera, M. Hiroi, K. Maezawa, S. Mock, H. v. Löhneysen, A. Bruckl, K. Neumaier, and K. Andres, J. Magn. Magn. Mater. **177-181**, 277 (1998).

6) P. S. Riseborough, Adv. Phys. **49**, 257 (2000).

7) V. M. T. Thiede, T. Ebel, and W. Jeitschko, J. Mater. Chem. 8, 125 (1998).

8) Y. Muro, K. Motoya, Y. Saiga, and T. Takabatake, J. Phys. Soc. Jpn. **78**, 083707 (2009).

9) A. M. Strydom, Physica B **404**, 2981 (2009).

10) N. Nishioka, Y. Kawamura, T. Takesaka, R. Kobayashi, H. Kato, M. Matsumura, K. Kodama, K. Matsubayashi, and Y. Uwatoko, J. Phys. Soc. Jpn. 78, 123705 (2009).

11) S. C. Chen and C. S. Lue, Phys. Rev. B **81**, 075113 (2010).

12) M. Matsumura, Y. Kawamura, S. Edamoto, T. Takesaka, H. Kato, T. Nishioka, Y. Tokunaga, S. Kambe, and H. Yasuoka, J. Phys. Soc. Jpn. **78**, 123713 (2009).

13) H. Tanida, D. Tanaka, M. Sera, C. Moriyoshi, Y. Kuroiwa, T. Takesata, T. Nishioka, H. Kato, and M. Matsumura, J. Phys. Soc. Jpn. **79**, 043708 (2010).

14) K. Hanzawa, J. Phys. Soc. Jpn. **79**, 043710 (2010).

15) I. Ishii, Y. Suemori, T.K. Fujita, T. Takesaka, T. Nishioka, and T. Suzuki, J. Phys. Soc. Jpn. **79**, 053602 (2010).

16) S. Kambe, H. Chudo, Y. Tokunaga, T. Koyama, H. Sakai, T. U. Ito, K. Ninomiya, W. Higemoto, T. Takesaka, T. Nishioka, and Y. Miyake, J. Phys. Soc. Jpn. **79**, 053708 (2010).

17) H. Tanida, D. Tanaka, M. Sera, C. Moriyoshi, Y. Kuroiwa, T. Takesata, T. Nishioka, H. Kato, and M. Matsumura, J. Phys. Soc. Jpn. **79**, 063709 (2010).

18) J. Robert, J.-M. Mignot, G. Andre, T. Nishioka, R. Kobayashi, M. Matsumura, H. Tanida, D. Tanaka, and M. Sera, Phys. Rev. B **82**, 100404(R) (2010).

19) Y. Muro, J. Kajino, K. Umeo, K. Nishimoto, R. Tamura, and T. Takabatake, Phys. Rev. B **81**, 214401 (2010).

20) K. Hanzawa, J. Phys. Soc. Jpn. **79**, 084704 (2010).

21) A. Kondo, J. Wang, K. Kindo, T. Takesaka, Y. Kawamura, T. Nishioka, D. Tanaka, H. Tanida, and M. Sera, J. Phys. Soc. Jpn. **79**, 073709 (2010).

22) H. Tanida, D. Tanaka, M. Sera, C. Moriyoshi, Y. Kuroiwa, T. Takesaka, T. Nishioka, H. Kato, and M. Matsumura, J. Phys. Soc. Jpn. **79**, 083701 (2010).

23) Y. Kawamura, S. Edamoto, T. Takesaka, T. Nishioka, H. Kato, M. Matsumura, Y. Tokunaga, S. Kambe, and H. Yasuoka, J. Phys. Soc. Jpn. **79**, 103701 (2010).

24) S.-i. Kimura, T. Iizuka, H. Miyazaki, A. Irizawa, Y. Muro, and T. Takabatake, Phys. Rev. Lett. **106**, 056404 (2011).

25) D. T. Adroja, A. D. Hillier, P. P. Deen, A. M. Strydom, Y. Muro, J. Kajino, W. A. Kockelmann, T. Takabatake, V. K. Anand, J. R. Stewart, and J. Taylor, Phys. Rev. B **82**, 104405 (2010).

26) D. D. Khalyavin, A. D. Hillier, D. T. Adroja, A. M. Strydom, P. Manuel, L. C. Chapon, P. Peratheepan, K. Knight, P. Deen, C. Ritter, Y. Muro, and T. Takabatake, Phys. Rev. B **82**, 100405(R) (2010).

27) A. Kondo, J. Wang, K. Kindo, T. Takesaka, Y. Ogane, Y. Kawamura, T. Nishioka, H. Tanida, and M. Sera, J. Phys. Soc. Jpn. **80**, 013701 (2011).

28) C. S. Lue, S. H. Yang, T. H. Su, and B. L. Young, Phys. Rev. B **82**, 195129 (2010).

29) H. Kato, R. Kobayashi, T. Takesaka, T. Nishioka, M. Matsumura, K. Kaneko, and N. Metoki, J. Phys. Soc. Jpn. **80**, 073701 (2011).

30) S. Kimura, Y. Muro, and T. Takabatake, J. Phys. Soc. Jpn. **80**, 033702 (2011).

31) P. P. Deen, D. T. Adroja, A. M. Strydom, A. D. Hillier, Y. Muro, J. Kajino, F. Demmell, P. Peratheepan, T. Takabatake, J. Taylor, and J. R. Stewart, private communication.

32) K. Umeo, T. Ohsuka, Y. Muro, J. Kajino, and T. Takabatake, J. Phys. Soc. Jpn. **80**, 064709 (2011).

33) A. Kondo, J. Wang, K. Kindo, Y. Ogane, Y. Kawamura, S. Tanimoto, T. Nishioka, D. Tanaka, H. Tanida, and M. Sera, Phys. Rev. B **83**, 180415(R) (2011).

34) M. Sakoda, S. Tanaka, E. Matsuoka, H. Sugawara, H. Harima, F. Honda, R. Settai, Y. Onuki, T. D. Matsuda, and Y. Haga, J. Phys. Soc. Jpn. **80**, 084716 (2011).

35) H. Tanida, D. Tanaka, M. Sera, S. Tanimoto, T. Nishioka, M. Matsumura, M. Ogawa, C. Moriyoshi, Y. Kuroiwa, J. E. Kim, N. Tsuji, and M. Tanaka, Phys. Rev. B **84**, 115128 (2011).

36) S. Kimura, T. Iizuka, H. Miyazaki, T. Hajiri, M. Matsunami, T. Mori, A. Irizawa, Y. Muro, J. Kajino, and T. Takabatake, Phys. Rev. B **84**, 165125 (2011).

37) H. Tanida, D. Tanaka, Y. Nonaka, M. Sera, M. Matsumura, and T. Nishioka, Phys. Rev. B **84**, 233202 (2011).

38) C.S. Lue and H.F. Liu, Phys. Rev. B **85**, 245116 (2012).

39) H. Tanida, Y. Nonaka, D. Tanaka, M. Sera, Y. Kuroiwa, Y. Uwatoko, T. Nishioka, and M. Matsumura, Phys. Rev. B **85**, 205208 (2012).

40) J. Goraus and A. Slebarski, J. Phys.: Condens. Matter **24**, 095503 (2012).

41) F. Stigari, T. Willers, Y. Muro, K. Yutani, T. Takabatake, Z. Hu, Y.-Y. Chin, S. Agrestini, H.-J. Lin, C. T. Chen, A. Tanaka, M. W. Haverkort, L. H. Tjeng, and A. Severing, Phys. Rev. B **86**, 081105 (2012).

42) K. Kunimori, M. Nakamura, H. Nohara, H. Tanida, M. Sera, T. Nishioka, and M. Matsumura, Phys. Rev. B **86**, 245106 (2012).

43) J. Robert, J.-M. Mignot, S. Petit, P. Steffens, T. Nishioka, R. Kobayashi, M. Matsumura, H. Tanida, D. Tanaka, and M. Sera, Phys. Rev. Lett. **109**, 267208 (2012).

44) M. Samsel-Czekala, E. Talik, M. Pasturel, and R. Troc, J. All. Com. **554**, 438 (2013).

45) M. Matsumura, N. Tomita, S. Tanimoto, Y. Kawamura, R. Kobayashi, H. Kato, T. Nishioka, H. Tanida, and M. Sera, J. Phys. Soc. Jpn. **82**, 023702 (2013).

46) F. Stigari, T. Willers, Y. Muro, K. Yutani, T. Takabatake, Z. Hu, S. Agrestini, Y.-Y.

Chin, H.-J. Lin, T. W. Pi, C. T. Chen, E. Weschke, E. Schierle, A. Tanaka, M. W. Haverkort, L. H. Tjeng, and A. Severing, Phys. Rev. B **87**, 125119 (2013).

47) A. Kondo, K. Kindo, K. Kunimori, H. Nohara, H. Tanida, M. Sera, R. Kobayashi, T. Nishioka, and M. Matsumura, J. Phys. Soc. Jpn. **82**, 054709 (2013).

48) D. T. Adroja, A. D. Hillier, Y. Muro, J. Kajino, T. Takabatake, P. Peratheepan, A. M. Strydom, P. P. Deen, F. Demmel, J. R. Stewart, J. W. Taylor, R. I. Smith, S. Ramos, and M. A. Adams, Phys. Rev. B **87**, 224415 (2013).

49) D. D. Khalyavin, D. T. Adroja, P. Manuel, J. Kawabata, K. Umeo, T. Takabatake, and A. M. Strydom, Phys. Rev. B **88**, 060403 (2013).

50) H. Guo, H. Tanida, R. Kobayashi, I. Kawasaki, M. Sera, T. Nishioka, M. Matsumura, I. Watanabe, and Z. A. Xu, Phys. Rev. B **88**, 115206 (2013).

51) D. T. Adroja, A.D. Hillier, Y. Muro, T. Takabatake, A.M. Strydom, A. Bhattacharyya, A. D. Aladin, and J. W. Taylor, Phys. Scr. **88**, 068505 (2013).

52) J. Kawabata, T. Takabatake, K. Umeo, and Y. Muro, Phys. Rev. B **89**, 094404 (2014).

53) O. Gunnarsson and K. Schönhammer, Phys. Rev. Lett. **50**, 604 (1983).

54) A. Sekiyama, S. Suga, T. Iwasaki, S. Uchida, S. Imada, Y. Saitoh, T. Yoshino, D. T. Adroja, and T. Takabatake, J. Elec. Spec. Relat. Phenom. **114-116**, 699 (2001).

55) J. J. Yeh and I. Lindau, At. Data Nucl. Data Tables **32**, 1 (1985).

56) H. Kumigashira, T. Takahashi, S. Yoshii, and M. Kasaya, Phys. Rev. Lett. **87**, 067206 (2001).

57) R. A. Rayjada, A. Chainani, M. Matsunami, M. Taguchi, S. Tsuda, T. Yokoya, S. Shin, H. Sugawara, and H. Sato, J. Phys.: Condens. Matter **22**, 095502 (2010).

58) M. J. Rozenberg, G. Kotliar, and H. Kajueter, Phys. Rev. B **54**, 8452 (1996).

Figure captions

Fig. 1 (color online) On-resonance (a) and off-resonance (b) spectra of Ce$M_2$Al$_{10}$ ($M$ = Fe, Ru, and Os) across the 3$d$-4$f$ threshold. Photon energies used are indicated by thick bars in the XAS spectra (c). In (a), an off-resonance spectrum of CeFe$_2$Al$_{10}$, which is normalized with scan numbers and incident photon flux, is also plotted to demonstrate the resonant enhancement (thin line).

Fig. 2 (color online)  (a)-(c) Temperature-dependent HRPES spectra near $E_F$ of $CeM_2Al_{10}$ ($M$ = Fe, Ru, and Os), respectively. The inset shows an enlargement near the $E_F$ region. (d)-(f) Normalized intensity at $E_F$ and (g)-(i) temperature-dependent normalized HRPES spectral intensity of $CeM_2Al_{10}$ ($M$ = Fe, Ru, and Os, respectively) deduced from (a)-(c). In (d)-(f), normalization was performed with the intensity at 70 K. The broken curve in (d) is a result of the polynomial fitting of the data from 10 to 120 K of $CeFe_2Al_{10}$ . The curves in (e) and (f) are the same as that in (d), but plotted with respect to the normalized temperatures using different $T_K$s, which are determined from 3 x $T_{max}$. $T_{max}$ is the temperature corresponding to the maximum magnetic susceptibility. In (g)-(i), the thick gray lines are visual guides representing the energy regions where the suppression of DOS starts. The red bars denote the shoulder structures.

Fig. 3 (color online)  Comparison of the normalized HRPES spectra near $E_F$ of $CeM_2Al_{10}$ ($M$ = Fe, Ru, and Os) measured at 10 K. The dotted lines are the fitting results for the higher-, middle-, and lower-binding-energy regions to determine the characteristic energies. The inset shows a schematic diagram of the changes in band dispersions (left) and partial DOSs (right) obtained using the *c-f* hybridization model.[6,56,58]

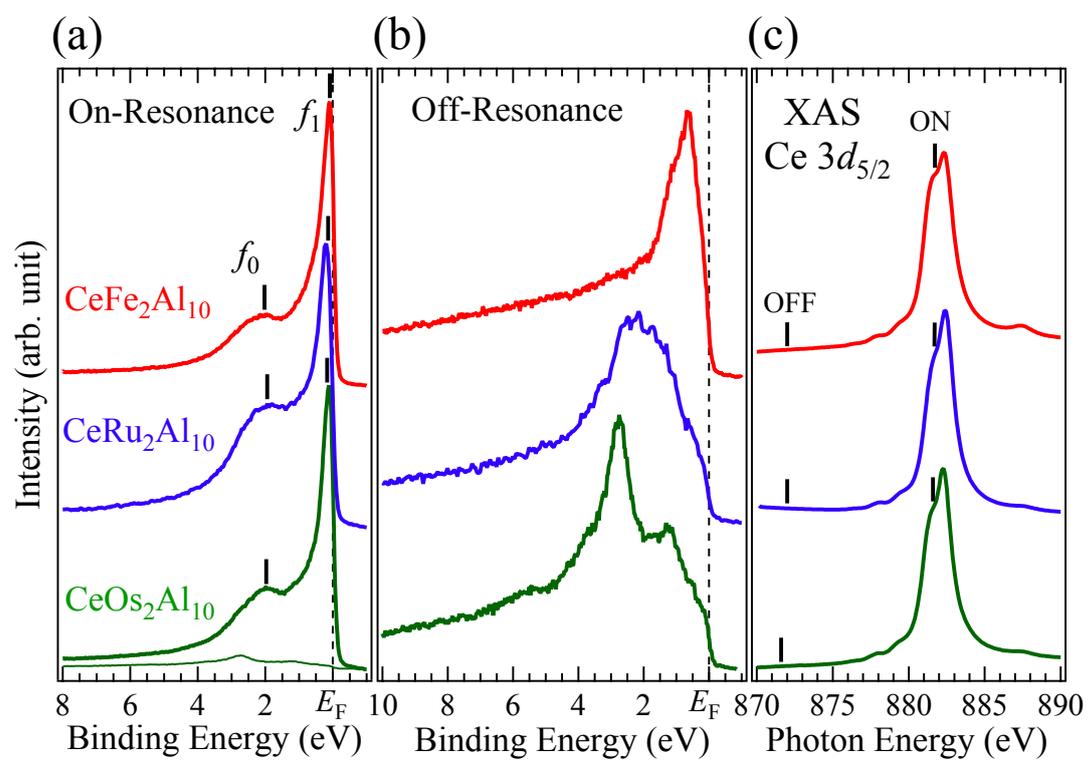

Fig. 1   T. Ishiga.

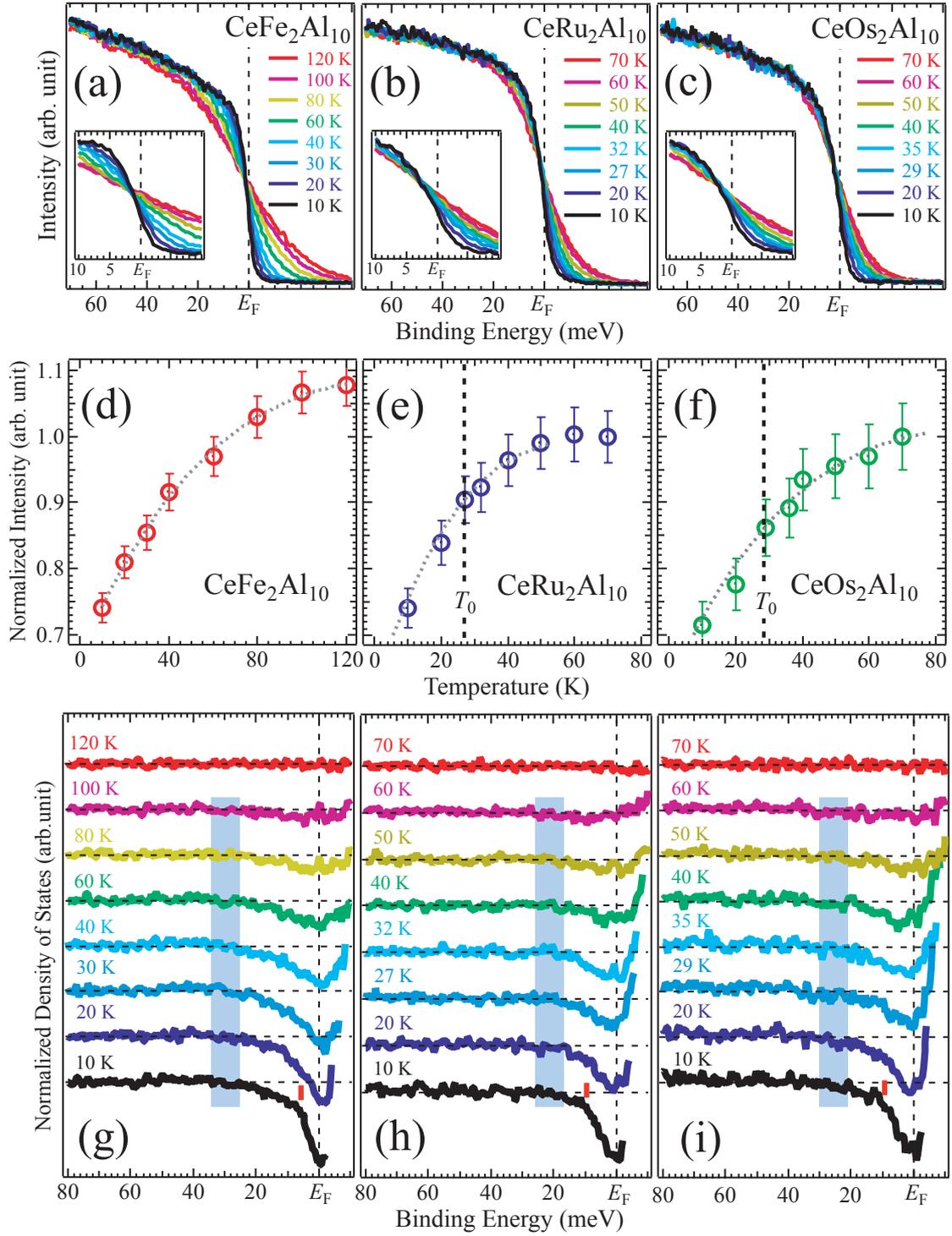

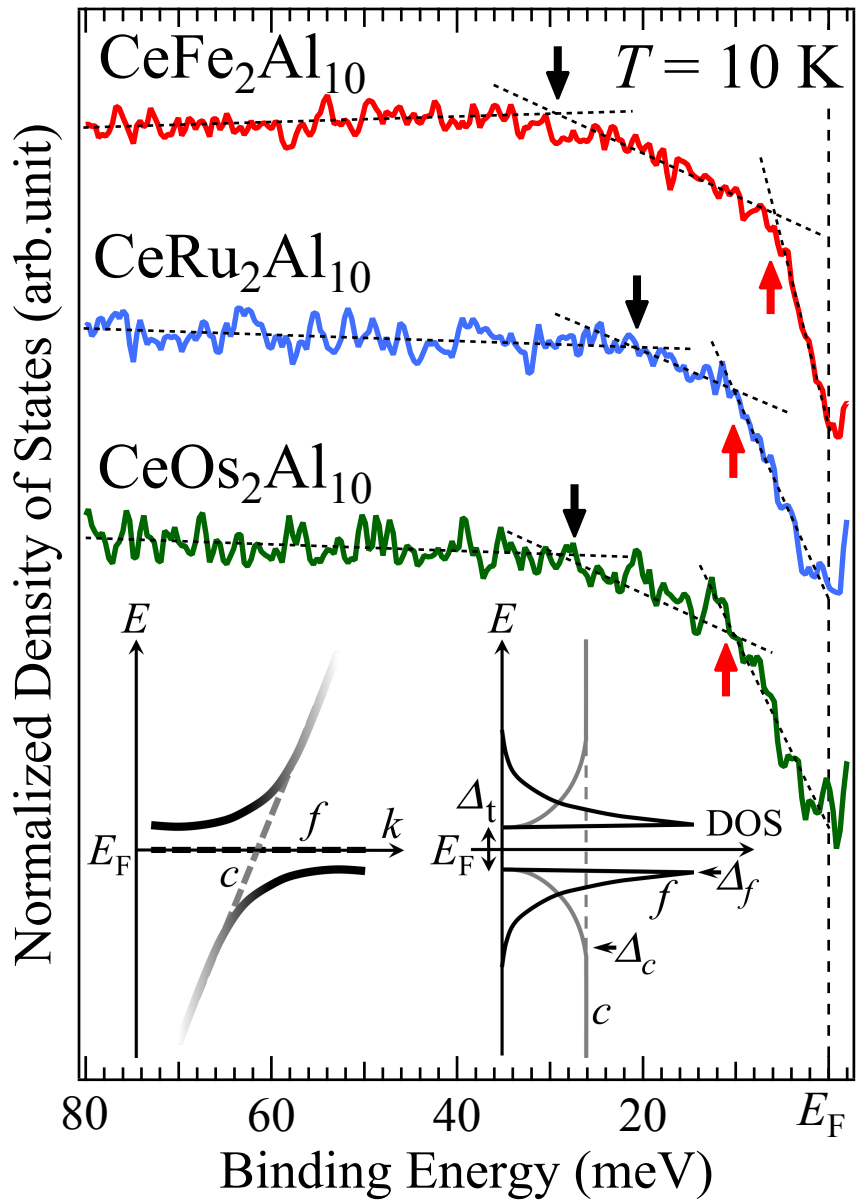

Fig. 3　T. Ishiga

17